# Flexibility Options: A Proposed Product for Managing Imbalance Risk

Elina Spyrou, *Member, IEEE,* Qiwei Zhang, *Member, IEEE,* Robin B. Hytowitz,
Benjamin F. Hobbs, *Life Fellow, IEEE,* Siddharth Tyagi, Mengmeng Cai, *Member, IEEE,* and Michael Blonsky, *Member, IEEE*

*Abstract*—The presence of variable renewable energy resources with uncertain outputs in day-ahead electricity markets results in additional balancing needs in real-time. Addressing those needs cost-effectively and reliably within a competitive market with unbundled products is challenging as both the demand for and the availability of flexibility depends on day-ahead energy schedules. Existing approaches for reserve procurement usually rely either on oversimplified demand curves that do not consider how system conditions that particular day affect the value of flexibility, or on bilateral trading of hedging instruments that are not co-optimized with day-ahead schedules. This article proposes a new product, 'Flexibility Options', to address these two limitations. The demand for this product is endogenously determined in the day-ahead market and it is met cost-effectively by considering real-time supply curves for product providers, which are co-optimized with the energy supply. As we illustrate with numerical examples and mathematical analysis, the product addresses the hedging needs of participants with imbalances cost-effectively, provides a less intermittent revenue stream for participants with flexible outputs, promotes value-driven pricing of flexibility, and ensures that the system operator is revenue-neutral. This article provides a comprehensive design that can be further tested and applied in large-scale systems.

## NOMENCLATURE

**Sets and Indices**

| | |
|---|---|
| $G$ | Set of market participants, indexed $i$ |
| $G^B$ | Set of FO buyers, subset of $G$ |
| $G^S$ | Set of FO sellers, subset of $G$ |
| $k$ | Index used for constraints |
| $R$ | Set of tiers, indexed $r$. $R = \{1, \ldots, r, \ldots, |S|-1\}$ |
| $S$ | Set of scenarios, indexed $s$, in ascending order wrt electricity generation. That is, $s=1$ corresponds to the lowest electricity generation of a FO buyer. |

This work was supported in part by funding provided by the U.S. Department of Energy Advanced Research Projects Agency–Energy (ARPA-E) under ARPA-E Award No. DE-AR001274 and Work Authorization No. 19/CJ000/07/08 and in part by a Leverhulme International Professorship, grant reference LIP-2020-002. E. Spyrou was with the the Power Systems Engineering Center of the National Renewable Energy Laboratory, Golden, CO 80401 USA. She is now with Imperial College London, UK (e-mail: elina.spyrou@ic.ac.uk). Q. Zhang was with the Johns Hopkins University, Baltimore, MD 21218 USA. He is now with the Grid Planning and Analysis Center, National Renewable Energy Laboratory, Golden, CO 80401 USA. R. B. Hytowitz was with the Electric Power Research Institute, Palo Alto, CA 94304 USA. She is now with NextEra Analytics, St Paul, MN 55107 USA. B.F. Hobbs is with the Johns Hopkins University, Baltimore, MD 21218 USA. S. Tyagi was with the Johns Hopkins University, Baltimore, MD 21218 USA. He is now with the Independent Electricity System Operator, Mississauga, ON L5J 4R9, Canada. M. Cai was with the the Power Systems Engineering Center of the National Renewable Energy Laboratory, Golden, CO 80401 USA. She is now with the Southeast University, China. M. Blonsky is with the Grid Planning and Analysis Center, National Renewable Energy Laboratory Golden, Colorado 80401 USA.

| | |
|---|---|
| $T$ | Set of time intervals, indexed $t$ |

**Day-ahead Parameters**

| | |
|---|---|
| $C_{i,t}^{\uparrow\|\downarrow}$ | RT strike price for FO up/down by seller $i \in G^S$ at time $t$ ($/MWh) |
| $\Pi_s$ | Probability of scenario $s$ |
| $P_{i,t}^{max}$ | Generator capacity of participant $i$ at $t$ (MW) |
| $P_{i,t}^{min}$ | Minimum generation when participant $i$ on at $t$ (MW) |
| $\Pi_r^{\uparrow\|\downarrow}$ | Probability of exercising FO up/down of tier $r$ |
| $VC_{i,t}^{\uparrow\|\downarrow}$ | Up/down FO scarcity cost for FO buyer $i \in G^B$ at $t$ |
| $C_{i,t}$ | Variable cost for participant $i$ at time $t$ ($ / MWh) |
| $D1/D2$ | Coefficients for penalty term in demand-supply balance |
| $D_t$ | DA inelastic demand at time $t$ (MW) |
| $M$ | Penalty term for $y_{i,s,t}$ ($/MW$) |
| $P_{i,s,t}$ | RT upper operating limit for FO buyer $i \in G^B$ at $t$ in scenario $s$ (MWh) |
| $RR_{i,t}$ | Ramp rate of FO supplier $i$ at $t$ (MW/hour) |

**Decision variables for DA problem**

| | |
|---|---|
| $hd_{i,r,t}^{\uparrow\|\downarrow}$ | FO up/down at tier $r$, time $t$ bought by $i \in G^B$(MW) |
| $hs_{i,r,t}^{\uparrow\|\downarrow}$ | FO up/down at tier $r$, time $t$ sold by $i \in G^S$(MW) |
| $sd_{i,r,t}^{\uparrow\|\downarrow}$ | Self-hedged up/down imbalance at tier $r$ for buyer $i \in G^B$ (MW) |
| $d^{DA}$ | Unmet DA load |
| $d_s^{RT}$ | Incremental to $d^{DA}$ unmet load in RT |
| $p_{i,t}^{DA/RT}$ | Participant's $i$ DA/RT electricity sales at $t$ (in MWh) |
| $u_{i,t}$ | Unit status of generator $i$ at $t$; $u_{i,t} \in \{0,1\}$ |
| $y_{i,s,t}$ | Auxiliary variable for absolute FO volume hedging FO buyer $i \in G^B$, in scenario $s$ at $t$ (MW) |

**Dual decision variables**

| | |
|---|---|
| $\lambda_t^{DA/RT}$ | DA/RT energy price at interval $t$ ($/ MWh) |
| $\lambda_{r,t}^{\uparrow\|\downarrow}$ | DA FO up/down price of tier $r$ at time $t$ (in $/ MW) |

**Real-time Parameters**

| | |
|---|---|
| $\alpha_{r,t}^{\uparrow\|\downarrow}$ | Portion of DA traded FOs in tier $r$ exercised in RT |
| $\overline{HD}_{i,r,t}^{\uparrow\|\downarrow}$ | FO volume FO buyer $i \in G^B$ can exercise in tier $r$ at $t$ based on DA-RT imbalance (MW) |
| $\overline{HS}_{r,t}^{\uparrow\|\downarrow}$ | FOs up/down in tier $r$ with strike prices lower/higher than the RT price (MW) |
| $VC_{r,t}^{\uparrow\|\downarrow}$ | RT strike price for FO up/down for tier $r$ at time $t$ (in $/ MWh) |
| $P_{i,t}^{RT}$ | RT available capacity for generator $i \in G^B$ (in MW) |

## I. INTRODUCTION

THE expansion of variable renewable energy resources (VRE) has impacted short-term (i.e., day-ahead (DA) to real-time (RT)) balancing needs. Due to VRE forecast errors, power system operators, electricity consumers and producers must manage imbalances more frequently and at higher magnitudes [1]. For instance, the California Independent System Operator (CAISO) and the Midcontinent Independent System Operator (MISO) have both experienced DA to RT imbalances up to 3 or 4 GW, a magnitude equivalent to multiple nuclear units. System operators have realized that managing imbalances in RT is at best expensive and at worst can threaten RT resource adequacy because RT rescheduling is limited by and depends on DA schedules. That is why many solutions are proposed both in the literature and in practice to develop flexible DA resource offers and schedules that anticipate the need for rescheduling to address DA-RT imbalances.

The spectrum of proposals range from modifying market inputs & bidding strategies (which we call Type I solutions), to fundamental redesign of DA markets (Type II), to introduction of new products with particular financial and/or physical features (Type III). Type I solutions are appealing because they can be immediately adopted, if they are not already followed in practice. There are several variants of Type I solutions. One subset follows common practice [2] by using virtual bids or financial derivatives to address participants' hedging needs [3]–[5]. This subset of solutions does not necessarily result in a cost-effective and reliable schedule of physical flexibility [5]. Another subset of Type I solutions modifies DA inputs (e.g., DA renewable generation [6], load forecasts [7], or reserve requirements [8]) intending to improve the cost-effectiveness and/or reliability of the physical schedule. The modifications can, however, be arbitrary when a DA price signal for flexibility is missing. Furthermore, their effectiveness heavily depends on the quality of the system operator's assumptions on RT imbalances and the costs of RT supply.

Type II solutions usually rely on Stochastic Unit Commitment (SUC) [9]. SUC, in theory, yields a cost-effective DA schedule, but it has two fundamental limitations. First, its dual variables cannot be directly used for DA prices [10] because they do not guarantee cost recovery per scenario. Several settlement schemes for SUC address this weakness at the expense of other objectives of market design, such as price convergence, revenue neutrality, cost-effectiveness [11]. Second is the curse of dimensionality that leads to highly simplified stochastic models that consider only a few possible scenarios and decision stages, with inevitably limited ability to find DA schedules with the hoped for efficiencies [12].

Last, Type III solutions are most actively pursued because they provide a clear price signal for flexibility and can facilitate transparent risk transfer. One set of Type III solutions include contracts and products that are centrally [13] or bilaterally [14] traded, but their procurement is not co-optimized with that of other DA market products. Another set of Type III solutions include new products that system operators procure within the DA market. Examples include ramping products or operating reserves in the USA [15]).

The former set of Type III solutions facilitate risk sharing among participants (e.g., wind power plants in [16]). A market party facing an imbalance risk pays a premium to another party who agrees to take on that risk by hedging it physically ([13], [14]) or financially [17]. 'Dual-trigger' options, such as the ones in [17], are particularly interesting because they are exercised upon the satisfaction of two conditions (e.g., (1) option buyer has an energy deficit; and (2) the imbalance price is higher than the pre-agreed strike price). This allows the market to benefit from discovering the optimal rescheduling in RT and providing the hedge to the party most needing it. Unfortunately, these option products have two major weaknesses. First, the lack of co-optimized procurement of those options with other market products complicates their pricing and the participation of sellers/buyers in other markets. As a result, their opportunity costs are uncertain because the buyer and seller are both unsure of prices and costs of other products, congestion, and other relevant market conditions that arise when scheduling DA energy and reserves. Second, existing literature has narrowly focused on call options that hedge only shortfalls in RT supply. Even for small systems, we find that baskets of call and put options, which hedge both deficits and surpluses of energy, are necessary for cost-effective hedging of DA-RT imbalances (see Section V).

The latter set of Type III solutions has been implemented in US ISO-managed markets. Co-optimizing procurement of reserves and other DA products facilitates joint price discovery and recognizes the opportunity costs that reserve suppliers face. However, the demand curve for those reserves is not linked with the hedging needs of resources with imbalances and is only partially linked with the RT balancing costs of flexible resources (allowing for availability payments and opportunity costs, but omitting deployment costs). These missing links challenge the cost-effectiveness of reserve products, which is shown to be very sensitive to the valuation assumed [18], [19]. Endogenous reserves, proposed in [20], fills some of this gap by considering balancing costs of reserve suppliers and by including resources with imbalances as reserve suppliers. However, price formation under endogenous reserves is not discussed in [20]; and it will likely face some of the same pricing challenges as SUC, given that the ISO will need to exogenously determine scenarios for imbalances. Ref. [21] argues that obtaining the optimal valuation of reserves is as elusive as the adoption of SUC. SUC is indeed used in [22] as a framework to construct demand curves for operating reserves, but the approach strongly assumes that the system operator can solve a stochastic reliability model or can optimally bias the load forecast.

The imbalance reserve (IR) product proposed by CAISO is an example of a Type III solution. The IR is a DA product that procures flexibility to manage DA-RT imbalances. In designing it, CAISO staff recognized that a link with the RT balancing costs of flexible resources is crucial and proposed a soft link through a cap for RT bid costs of IR suppliers. The cap proposal was abandoned because its calculation during tight supply conditions proved complicated [23]–[26]. The CAISO's proposal also recognized the need for departing from over-simplified settlement schemes for ancillary services [27] and charges IR procurement costs to resources that are responsible for RT-DA imbalances. This cost causation link is ex-post and differs from the ex-ante link in [28] that charges DA uncertainty-driven costs to resources contributing to *anticipated extreme* imbalances. By allocating reserve costs to entities contributing to the need for those reserves, they are motivated to reduce those needs. A challenge however is ensuring revenue adequacy; for the IR, CAISO's allocation is based on *realized* imbalances and charges resources no more than the DA IR price. Given that *realized* imbalances are frequently lower than the *extreme* imbalance considered in IR requirements, the cost allocated to resources could be less than the revenue IR suppliers receive. Hence, the operator will likely not meet the goal of revenue adequacy, and will have to recover shortfalls from other market parties [26].

This article proposes a novel dual-trigger option, 'Flexibility Option' (FO), which advances the state-of-the-art in Type III solutions for the management of short-term net load uncertainty. The product's novel characteristics include:

- determining the demand for flexibility, *within the DA market*, based on the probability distribution of imbalances in relation to the co-optimized DA energy schedule of FO buyers;
- consideration of balancing costs, again *within the DA market*, using RT strike prices submitted by FO sellers;
- multiple classes ('tiers') of FOs with different probability of exercise and corresponding costs, rather than a single reliability target, such as the CAISO IR 95% goal; and
- a revenue-adequate settlement scheme that follows cost causation principles.

Existing and proposed reserve products rely on an exogenous single (ISO) view of RT market outcomes. However, in our design, FO buyers express their individual hedging needs by submitting a discrete probability distribution of their RT output. Similarly, FO sellers reveal their balancing costs (i.e., how much it would cost them to hedge imbalances) by financially committing to RT strike prices. Therefore, in addition to opportunity costs within the DA market, the pricing of FOs accounts for trade opportunities in RT. Not only does this yield more accurate pricing of FOs compared to traditional reserves, it also enables resources to buy hedges based on their imbalance risk while avoiding allocating costs to resources who don't need hedges. The co-optimization also yields dynamic FO prices, because it endogenously considers the trade-off between reserving flexibility to back up FO sales and selling other products. Such dynamic pricing would be more difficult, if not impossible, in private bilateral markets.

In addition to the FO conceptual design in Section II and specific mathematical formulations in Section III, Section IV presents a numerical study designed to inform stakeholders about FOs' strengths and weaknesses relative to traditional reserve products. The results in Section V demonstrate: (a) how the FO design promotes cost-effectiveness, value-driven pricing of flexibility, and revenue adequacy for the system operator; and (b) how the FO sellers and buyers experience less intermittent revenues for flexibility and hedge their volatile costs for imbalances, respectively. We summarize our findings in Section VI and provide examples and mathematical analysis for selected properties of the FO design in the Appendices.

## II. Flexibility Options: Preliminaries

We propose the introduction of Flexibility Options as a class of products with voluntary participation[1] in systems where operators want to efficiently manage imbalances between at least two electricity markets, including DA and RT.

FOs are dual-trigger options in which the purchaser obtains the right to buy or sell at a strike price $VC^{\uparrow|\downarrow}$ RT energy imbalances with respect to a DA trigger quantity $Q^{Trigger}$. Similar to operating reserves, we define two FO types: upward and downward. Note that here and in the rest of the paper, we use the notation $x^{\uparrow|\downarrow}$ as a shorthand, indicating that a variable or parameter x actually corresponds to two variables/parameters: $x^{\uparrow}$ and $x^{\downarrow}$, the first associated with the up direction (the supplier increases output if the option is exercised) and the other with the down direction. The two FO types differ with respect to the direction of inequality for the trigger conditions. Upward FOs are exercised when both (1a) and (1b) hold:

$$P_t^{RT} < Q_t^{Trigger,\uparrow} \quad (1a) \qquad \lambda_t^{RT} > VC_t^{\uparrow} \quad (1b)$$

Downward FOs are exercised when both (2a) and (2b) hold:

$$P_t^{RT} > Q_t^{Trigger,\downarrow} \quad (2a) \qquad \lambda_t^{RT} < VC_t^{\downarrow} \quad (2b)$$

Eqs. (1a) and (2a) are written in a general form that would allow their application either individually for each FO buyer or for a group of FO buyers. In case the conditions are met simultaneously for all FO buyers, this distinction does not matter. In practice, the system operator can determine how to apply the scheme as long as strike-up prices are higher than strike-down prices.

To illustrate how the dual-trigger option works, consider a simple case with one resource that regularly experiences imbalances and another resource that can ramp up and down in RT. The former resource (hereafter called *uncertain*) would be a FO buyer and the latter resource (hereafter called *flexible*) would be a FO seller. If FOs do not exist, the uncertain resource sells electricity $p_{i,t}^{DA}$ in DA for delivery at hour $t$. Then, in RT it must buy at least (or sell up to) $p_{i,t}^{DA} - P_{i,t}^{RT}$ at the RT price $\lambda_t^{RT}$ when its maximum RT operating point $P_{i,t}^{RT}$ is lower (or higher) than $p_{i,t}^{DA}$, respectively. The resource's RT production level depends on whether its variable cost is less than or equal to the RT price. When the uncertain resource buys electricity in RT due to a production shortfall, its RT gross margin could be negative. Depending on its costs, the prices in the two markets, and the quantities it sells DA and buys in RT, the resource could potentially suffer losses on the net over the two (DA, RT) markets. When the uncertain resource curtails output in RT to avoid losses, it foregoes any output-related revenues (e.g., RT revenues plus production tax credits). On the other hand, in RT, the flexible resource will add to any profit margin it earned DA as it will ramp up or down when its cost is lower or higher than $\lambda_t^{RT}$, respectively. These improved profit margins are volatile and unpredictable.

To hedge against volatile and potentially negative RT gross margins, the uncertain resource can buy FOs in DA with the DA energy schedule as the trigger quantity. The settlement of FO work as follows. First, consider the case of an *upward* FO. If the RT upper operating limit falls short of the trigger quantity, the buyer of an upward FO can exercise the option to receive a payment of $max(0, \lambda_t^{RT} - VC_t^\uparrow)$ per MWh of imbalance (i.e., $max(0, Q_t^{Trigger,\uparrow} - P_{i,t}^{RT})$). That way, the FO buyer offsets its RT losses from having to buy from the RT market its shortfall, which is caused by its RT capability being lower than its DA schedule.

---

[1]The product could become mandatory similar to the credit collateral requirement ISOs impose to participants [29].

Second, consider the case of an *downward* FO. When the uncertain resource is able to produce more than its DA position in RT, it can exercise the *downward* FO to receive a payment in RT of $max(0, VC_t^\downarrow - \lambda_t^{RT})$ per MWh of its imbalance (i.e., $max(0, P_{i,t}^{RT} - Q_t^{Trigger,\downarrow})$). This option payment is on top of the gross margin it gets from selling its excess RT output to the market at price $\lambda_t^{RT}$. That way, the FO buyer improves its revenues in case of surpluses.

On the other side of the deal is a flexible resource that supplies the FO. To secure a revenue stream that is less volatile than the RT gross margins it would earn by just playing the RT market, the flexible resource can also sell FOs in the DA market. That way, the FO seller foregoes at least part of its variable RT gross profit margins to provide the RT payoff to the FO buyer in exchange for a less variable DA FO premium. This is true for upward as well as downward FOs. In either case, the seller potentially reduces its risk, as measured by the standard deviation of its overall DA and RT returns.

A risk-neutral uncertain resource would buy FOs if the FO premium it pays for the option in DA is lower or equal to the probability-weighted RT FO payoff it would receive. Similarly, a risk-neutral FO seller would accept a DA premium in exchange for selling the FO if that premium is greater or equal to the (probability-weighted) foregone RT gross profit margin (i.e., the FO's payoff). Hence, for a trade to be attractive to both risk-neutral parties, the DA FO premium should be equal to the probability-weighted FO RT payoff. Pricing the FOs in DA would require accurately projecting RT scenarios $s$ for $\{P_{i,t}^{RT}, \lambda_t^{RT}\}$ and their corresponding probabilities, and requesting sellers and buyers to agree on $VC^{\uparrow|\downarrow}$ and $Q^{trigger}$.

Instead of projecting scenarios, we propose that the system operator facilitate FO trading and pricing in the following way. Prior to the DA market gate closure, the system operator specifies the number ($|S|$) of trigger quantities ($P_{i,s,t}$) it will consider in the DA market clearing. We follow the convention that the lower the $s$, the lower the trigger quantity for an uncertain supplier of electricity. Hence, the lower the $s$ the lower the probability that an upward FO will be activated to hedge a shortfall with respect to that trigger quantity. The system operator considers production shortfalls with respect to each trigger quantity ($P_{i,s,t}$) other than that the lowest trigger quantity (i.e., $P_{i,s=1,t}$). Therefore, the system operator facilitates trading of $|R| = |S| - 1$ upward FOs and chooses $|R|$ distinct probabilities of shortfall ($\Pi_r^\uparrow$) that it will consider in the DA market clearing. Symmetrically, the system operator considers production surpluses with respect to each trigger quantity ($P_{i,s,t}$) other than the highest trigger quantity (i.e.,

$P_{i,s=|S|,t}$) and facilitates trading of $|R|$ downward FOs, choosing $|R|$ distinct probabilities of surplus ($\Pi_r^\downarrow$) it will consider in the DA market clearing.

FO buyers have to submit multiple ($|R| + 1$) trigger quantities, each corresponding to a pair of ISO-determined probabilities of production shortfall and surplus. In particular, the pairs of probabilities for $\{P_{i,s=1,t}, P_{i,s=2,t}, \cdots P_{i,s=|S|,t}\}$ are $\{(0, \Pi_{r=1}^\downarrow), (\Pi_{r=1}^\uparrow, \Pi_{r=2}^\downarrow), \cdots (\Pi_{r=|S|-1}^\uparrow, 0)\}$. FO buyers can choose the trigger quantities based on probabilistic forecasts that they procure from commercial vendors (e.g., UL [30]) or that they develop with in-house tools similar to ones used by some system operators [31]. FO sellers have to submit their own seller-specific strike prices ($C^{\uparrow|\downarrow}$).

To keep the problem computationally tractable, the system operator can choose a relatively small value for $|R|$. Using the trigger quantities submitted by FO buyers, the system operator can construct $|R| + 1$ scenarios for the RT output of each FO buyer and ensure that each FO buyer is hedged.

The FO buyer is hedged with a basket of FOs that have different trigger quantities. Any upward FO would be hedging shortfalls of the FO buyer in RT when the FO's trigger quantity is greater or equal to the FO buyer's RT upper operating limit. Similarly, any downward FO would be hedging surpluses of the FO buyer in RT when the FO's trigger quantity is lower or equal to the FO buyer's RT upper operating limit. We refer to FO sub-types with different trigger quantities as *tiers*. In lay terms, tiers correspond to different qualities or degrees of risk hedging. For up or down FOs, their value increases when the trigger quantity is higher or lower, respectively. This is because conditions (1a) and (2a) are met more frequently for higher or lower trigger quantities, respectively. As shown in Fig. 1, the tiers $r$ of up FOs have trigger quantities $\{Q_1^\uparrow, \ldots, Q_{|R|}^\uparrow\}$ corresponding to $\{P_{i,s=2,t}, \ldots, P_{i,s=|R|+1,t}\}$, and the tiers of down FOs have trigger quantities $\{Q_1^\downarrow, \ldots, Q_{|R|}^\downarrow\}$ corresponding to $\{P_{i,s=1,t}, \ldots, P_{i,s=|R|,t}\}$. In the proposed design, the system operator assumes that up and down FOs in each tier are activated with probability $\Pi_{r=1\cdots|R|}^\uparrow$ and $\Pi_{r=1\cdots|R|}^\downarrow$, respectively. As an example, being short with respect to $P_{i,s=2,t}$ corresponds to only one case of RT upper operating limit (i.e., $P_{i,s=1,t}$). Hence, the probability of activating up-tier 1 ($\Pi_{r=1}^\uparrow$) corresponds to the cumulative probability of $P_{i,s=1,t}$. Similarly, being short with respect to $P_{i,s=3,t}$ includes all cases with RT upper operating limit lower or equal to $P_{i,s=2,t}$. Hence, the probability of activating up-tier 2 ($\Pi_{r=2}^\uparrow$) corresponds to the cumulative probability of $P_{i,s=2,t}$ and so on.

The system operator aggregates demand for FOs (e.g., by anticipating simultaneous exercise by all FO buyers of the FOs in each tier); it reserves ramping capability from FO sellers to meet the FO demand by solving a DA market that co-optimizes procurement of FOs and other products considering the probability-weighted strike prices as the balancing costs of FO suppliers. The dual variables of the FO supply-demand balance constraint provide a probability-weighted estimate of $\lambda_t^{RT}$ and are used to compensate FO sellers in DA. In RT, conditions (1b) and (2b) are checked at individual or aggregate FO buyer level and conditions (1a) and (2a) are checked for each FO seller to calculate $VC^{\uparrow|\downarrow}$ and FO payoffs. For more details and the precise mathematical formulation, please see Section III.

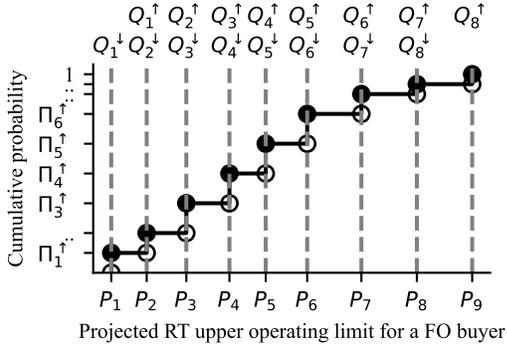

Fig. 1. Illustration of the discrete cumulative probability distribution for the RT upper operating limit ISO considers based on DA FO buyer inputs.

We choose to ask FO buyers to submit trigger quantities $P_{i,s,t}$, which are projections of their RT upper operating limit, for multiple reasons. First and foremost, when the demand for flexibility is explicitly introduced by resources with DA-RT imbalances, it is easy to set up settlement schemes that follow cost-causation principles. Second, the market monitor can use standard performance metrics of percentile forecasts such as reliability plots to assess how truthful/inaccurate the submissions of $P_{i,s,t}$ are over a period of time. By explicitly considering the probability of exercise of FOs, this design could better integrate flexibility from resources that can be flexible at some times but not others (e.g., retail electricity customers with priority of service contracts [32]).

## III. Formulation

To facilitate the trading of FOs, a system operator must modify their DA market formulation and settlement formulas according to sections III-A and III-B, respectively. While the introduction of FOs requires new settlement formulas in both DA and RT, the optimization model used to schedule generation in RT market does not itself change.

### A. Day-ahead Market Formulation

The DA market formulation ((3)-(13)) co-optimizes procurement of energy and FOs. Compared to the traditional DA unit commitment-based market formulation, it includes new decision variables for FOs, a new set of objective function terms for the probability-weighted cost increases/reductions caused by the exercise of FOs, and new constraints that endogenously determine the demand for FOs and ensure that FO suppliers have enough up and down headroom and ramping ability to provide FOs. The inclusion of the latter constraints makes the FO an engineering-economic product. Unlike a purely financial product, the FO reserves physical flexibility in resources to supply that product if called in RT.

$$\underset{\Xi}{\text{minimize}} \quad Objective^{DA}$$
$$\text{where} \quad \Xi = \{d_t^{DA}, d_{s,t}^{RT}, p_{i,t}^{DA}, hs_{i,r,t}^{\uparrow|\downarrow}, hd_{i,r,t}^{\uparrow|\downarrow} sd_{i,r,t}^{\uparrow|\downarrow}, u_{i,t}, y_{i,s,t}\} \tag{3}$$

$Objective^{DA}$ is defined as follows:

$$Objective^{DA} = \sum_{i \in G, t \in T} C_{i,t} \cdot p_{i,t}^{DA} \tag{4a}$$

$$\pm \sum_{i \in G^S, r \in R, t \in T} \Pi_{r,t}^{\uparrow|\downarrow} \cdot C_{i,t}^{\uparrow|\downarrow} \cdot hs_{i,r,t}^{\uparrow|\downarrow} \tag{4b}$$

$$\mp \sum_{i \in G^B, r \in R, t \in T} \left( \Pi_{r,t}^{\uparrow|\downarrow} \cdot C_{i,t} \cdot (hd_{i,r,t}^{\uparrow|\downarrow} + sd_{i,r,t}^{\uparrow|\downarrow}) \right) \tag{4c}$$

$$\pm \sum_{i \in G^B, r \in R, t \in T} \left( \Pi_{r,t}^{\uparrow|\downarrow} \cdot \widetilde{VC_{i,t}^{\uparrow|\downarrow}} \cdot sd_{i,r,t}^{\uparrow|\downarrow} \right) \tag{4d}$$

$$+ M \cdot \sum_{i,s,t} y_{i,s,t} \tag{4e}$$

$$+ \sum_{s} \Pi_s \cdot (D1 \cdot (d_t^{DA} + d_{s,t}^{RT}) + D2 \cdot (d_t^{DA} + d_{s,t}^{RT})^2) \tag{4f}$$

Note that $\pm$ above indicate that there is one term for FOs in the up direction that is added, and one term for FOs down that is subtracted ($\mp$ indicates the reverse). The above objective is minimized subject to the following constraints:

$$p_{i,t}^{DA}, hs_{i,r,t}^{\uparrow|\downarrow}, hd_{i,r,t}^{\uparrow|\downarrow}, sd_{i,r,t}^{\uparrow|\downarrow}, y_{i,s,t}, d_t^{DA} \geq 0$$

$$\sum_{i \in G} p_{i,t}^{DA} + d_t^{DA} = D_t \quad \forall t \in T \quad \text{(dual: } \lambda_t^{DA}) \tag{5}$$

$$\sum_{i \in G^S} hs_{i,r,t}^{\uparrow|\downarrow} = \sum_{i \in G^B} hd_{i,r,t}^{\uparrow|\downarrow} \quad \forall r \in R, \forall t \in T \quad \text{(dual: } \lambda_{r,t}^{\uparrow|\downarrow}) \tag{6}$$

**For FO buyers:** $\forall i \in G^B, \forall t \in T \quad (\forall r \in R, \text{ or } \forall s \in S)$

$$-d_{s,t}^{RT} + p_{i,t}^{DA} + \sum_{\substack{r \in \{1,\ldots,s-1\} \\ s \neq 1}} \left( hd_{i,r,t}^{\downarrow} + sd_{i,r,t}^{\downarrow} \right) - \sum_{\substack{r \in \{s,\ldots,|S|-1\} \\ s \neq |S|}} \left( hd_{i,r,t}^{\uparrow} + sd_{i,r,t}^{\uparrow} \right) = P_{i,s,t} \quad \text{(dual: } \lambda_{i,s,t}^{6}) \tag{7}$$

$$\sum_{\substack{r \in \{1,\ldots,s-1\} \\ s \neq 1}} \left( hd_{i,r,t}^{\downarrow} + sd_{i,r,t}^{\downarrow} \right) + \sum_{\substack{r \in \{s,\ldots,|S|-1\} \\ s \neq |S|}} \left( hd_{i,r,t}^{\uparrow} + sd_{i,r,t}^{\uparrow} \right) \leq y_{i,s,t} \quad \text{(dual: } \lambda_{i,s,t}^{7}) \tag{8}$$

$$y_{i,s,t} \geq p_{i,t}^{DA} - P_{i,s,t} \quad \text{(dual: } \lambda_{i,s,t}^{8}) \tag{9}$$

$$y_{i,s,t} \geq P_{i,s,t} - p_{i,t}^{DA} \quad \text{(dual: } \lambda_{i,s,t}^{9}) \tag{10}$$

**For FO sellers:** $\forall i \in G^S, \forall t \in T$

$$\sum_{r} hs_{i,r,t}^{\uparrow|\downarrow} \leq u_{i,t} \cdot RR_i \quad \text{(dual: } \lambda_{i,t}^{10}) \tag{11}$$

$$p_{i,t}^{DA} + \sum_{r} hs_{i,r,t}^{\uparrow} \leq P_{i,t}^{max} \cdot u_{i,t} \quad \text{(dual: } \lambda_{i,t}^{11}) \tag{12}$$

$$p_{i,t}^{DA} - \sum_{r} hs_{i,r,t}^{\downarrow} \geq P_{i,t}^{min} \cdot u_{i,t} \quad \text{(dual: } \lambda_{i,t}^{12}) \tag{13}$$

The objective function consists of the production costs for DA schedules (4a), the probability-weighted RT costs of FO suppliers (4b) and buyers (4c), the probability-weighted RT costs of resources that self-hedge their DA schedules (4d), a term that chooses among alternative optima (4e), and a quadratic penalty term (4f) that is needed to model the supply-demand balance as a soft constraint to avoid degeneracy.

In particular, in (4b) we assume that the FO suppliers incur additional costs at $C^{\uparrow}$ or save costs at $C^{\downarrow}$ when up or down FOs are exercised, respectively. Symmetrically, buyers of up/down FOs incur lower/higher production costs (compared to what is assumed in the DA market) when they experience deficits or surpluses (4c). Term (4d) is similar to term (4b), with FO buyers self-hedging their imbalances partially or fully ($sd_{i,r,t}^{\uparrow|\downarrow}$ reflects self-hedged MW) when the marginal strike prices for up or down FOs are higher or lower, respectively, than their FO scarcity cost ($\widetilde{VC^{\uparrow|\downarrow}}$).

Constraint (5) is the DA demand-supply energy balance. Constraint (6) balances supply and demand for up and down FOs. In this formulation, for each tier, the total FO demand is the sum of demand by all FO buyers, considering the possibility of simultaneous exercise by all FO buyers of the FOs they hold in each tier.

Given that the demand and supply of FOs depend on DA energy schedules, the formulation constraints the possible amounts of FOs considering the DA energy awards to FO buyers and sellers. For FO buyers, (7) ensures FO volume is adequate to manage imbalances between the forecasted RT potential of each FO buyer $P_{i,s,t}$ and their DA energy award $p_{i,t}^{DA}$. For each level of $P_{i,s,t}$, FOs up from tiers $r \geq s$ and FOs down from tiers $r \leq s-1$ hedge production imbalances.

For FO sellers, eqs. (11)-(13) ensure that the FOs sold are within the seller's technical capabilities, i.e., hourly ramping (11); generating capacity (12); and minimum generation level (13). Eqs. (11)-(13) do not allow *over-booking* of flexibility, i.e., selling more FOs than technically possible. Overbooking is commonly practiced in other industries (e.g., airlines that anticipate only a subset of ticketholders boarding a flight); and future research could explore its benefits for FO sellers.

Last, objective function term (4e) in conjunction with (8)-(10) help choose among alternative optima. It turns out that, in general, many different portfolios of FO could provide precisely the same risk hedging benefit (and RT schedule for FO suppliers) at the same cost. To facilitate interpretation and implementation of the solution, we add this term, which causes the model to choose the basket of FOs with the lowest absolute volume of FOs at any scenario $s$ from among those portfolios. We provide an example of alternative optima in Appendix D.

We intentionally omit variables and constraints related to other ancillary services, such as regulating reserves, as well as inter-temporal trajectories related to ramping, start-up, and shut down of units to focus the presentation on the novel features of the formulation.

### B. Settlements

Relative to traditional spot market settlements, settlements change both in DA and RT. The FO settlement scheme has two attractive qualities. First, it follows cost-causation principles; i.e., participants that cause the need for FOs pay the costs of FOs. Second, it keeps the system operator *revenue neutral*. In DA, charges from FO buyers are used to pay FO sellers; then in RT, charges to FO sellers result in cash flows to FO buyers.

*1) Day-ahead Settlements:* FO sellers ($i \in G^S$) receive a payment for each MW of FOs they sell at each $r, t$, i.e.,

$$\left(\lambda_{r,t}^{\uparrow|\downarrow} \mp \Pi_{r,t}^{\uparrow|\downarrow} \cdot C_{i,t}^{\uparrow|\downarrow}\right) \tag{14}$$

The FOs within each tier are valued at a single tier price in DA ($\lambda_{r,t}^{\uparrow|\downarrow}$). However, the net per MW premium FO sellers receive vary among sellers within a tier because FO sellers commit to different strike prices $C_{i,t}^{\uparrow|\downarrow}$.

On the other hand, FO buyers ($i \in G^B$) pay a uniform per MW price for FOs at each $r, t$. The uniform price is the weighted by quantity average of the total payments to FO sellers, i.e.,

$$\frac{\sum_{i \in G^S} \left(\lambda_{r,t}^{\uparrow|\downarrow} \mp \Pi_{r,t}^{\uparrow|\downarrow} \cdot C_{i,t}^{\uparrow|\downarrow}\right) \cdot hs_{i,r,t}^{\uparrow|\downarrow}}{\sum_{i \in G^B} hd_{i,r,t}^{\uparrow|\downarrow}} \tag{15}$$

*2) Real-time Settlements:* In RT, the settlements reverse in direction, i.e., the beneficiaries of DA credits are charged in RT, while the parties who incurred DA charges instead benefit from RT credits. We calculate RT settlements in three steps.

First, we estimate $\overline{HD}_{i,r,t}^{\uparrow}$ and $\overline{HD}_{i,r,t}^{\downarrow}$, i.e., the total MWh of up and down FOs that the FO buyers could exercise considering only one of the two triggers: the imbalances of $P_{i,t}^{RT}$ with respect to the FO trigger quantities. Equations (16)-(17) estimate the imbalance of a FO buyer in tier $r$ by comparing $P_{i,t}^{RT}$ to the tier's bounds $[P_{i,r,t}, P_{i,r+1,t}]$ and then limiting the RT demand for FOs to the DA FO purchase. For concise formulas, we use $\underline{M}(A, B) = min(A, B)$ and $t^+ = max(0, t)$. When the RT upper operating limit for an uncertain resource falls between two trigger quantities, the volume that the FO buyer can exercise is limited by the RT upper operating limit $P_{i,t}^{RT}$ and is possibly lower than the FO volume bought in the tier. That's why in equations (16)-(17), we have included $\underline{M}$.

$$\overline{HD}_{i,r,t}^{\uparrow} = \underline{M}(hd_{i,r,t}^{\uparrow}, [P_{i,r+1,t} - P_{i,t}^{RT}]^+) \tag{16}$$

$$\overline{HD}_{i,r,t}^{\downarrow} = \underline{M}(hd_{i,r,t}^{\downarrow}, [P_{i,t}^{RT} - P_{i,r,t}]^+) \tag{17}$$

Parameter $\alpha^{\uparrow|\downarrow}$ expresses the RT demand from FO buyers in relation to the FO volume purchased in DA.

$$\alpha_{r,t}^{\uparrow|\downarrow} = \frac{\sum_{i \in G^B} \overline{HD}_{i,r,t}^{\uparrow|\downarrow}}{\sum_{i \in G^B} hd_{i,r,t}^{\uparrow|\downarrow}} \tag{18}$$

Second, we compare the RT market-clearing energy price to FO sellers' strike prices (the $2^{nd}$ of two triggers considered) to find the RT FO supply per tier. For the case of FOs up, we identify the sellers of those FOs in the DA market whose own strike prices are lower than the RT price ($\overline{HS}_{r,t}^{\uparrow}$). The reverse is the case for FOs down; any sellers whose own strike prices exceed the RT price are considered in (19).

$$\overline{HS}_{r,t}^{\uparrow|\downarrow} = \sum_{\{i \in G^S | C_{i,t}^{\uparrow|\downarrow} <|> \lambda_t^{RT}\}} hs_{i,r,t}^{\uparrow|\downarrow} \tag{19}$$

To calculate the system-wide RT strike price for each tier, we activate a portion of the maximum possible volume of FOs that can be exercised ($\overline{HS}_{r,t}^{\uparrow|\downarrow} \cdot \alpha_{r,t}^{\uparrow|\downarrow}$) (see $1^{st}$ term in (20))[2] and any remaining RT FO demand is met at the RT price (see $2^{nd}$ term in (20)). In summary, the system-wide strike price in

$$VC_{r,t}^{\uparrow|\downarrow} = \frac{\sum\limits_{\{i \in G^S | C_{i,t}^{\uparrow|\downarrow} < | > \lambda_t^{RT}\}} C_{i,t}^{\uparrow|\downarrow} \cdot hs_{i,r,t}^{\uparrow|\downarrow} \cdot \alpha_{r,t}^{\uparrow|\downarrow}}{\sum_{i \in G^B} \overline{HD}_{i,r,t}^{\uparrow|\downarrow}} \quad (20)$$
$$+ \frac{[\left(\sum_{i \in G^B} \overline{HD}_{i,r,t}^{\uparrow|\downarrow} - \alpha_{r,t}^{\uparrow|\downarrow} \cdot \overline{HS}_{r,t}^{\uparrow|\downarrow}\right)]^+ \cdot \lambda_t^{RT}}{\sum_{i \in G^B} \overline{HD}_{i,r,t}^{\uparrow|\downarrow}}$$

---

[2] We multiply the RT FO supply by $\alpha_{r,t}^{\uparrow|\downarrow}$ to ensure that for each FO supplier, the probability-weighted volume exercised is equal to the probability-weighted volume sold. To be specific, the probability-weighted volume sold by a FO supplier $i$ in tier $r$ is $\Pi_{r,t}^{\uparrow} \cdot hs_{i,r,t}^{\uparrow}$. The FO formulation in DA assumes perfect correlation of imbalances among FO buyers. However, this is not necessarily true. As a result, it is possible that flexibility from an infra-marginal FO supplier would be activated in RT with a higher probability than $\Pi_{r,t}^{\uparrow}$ and the probability-weighted volume exercised will exceed the level sold. Assuming that each FO buyer's forecast is well calibrated (i.e., $E(\overline{HD}_{i,r,t}^{\uparrow}) = \Pi_{r,t}^{\uparrow} \cdot hd_{i,r,t}^{\uparrow}$), it is easy to show that $E(\alpha_{r,t}^{\uparrow} \cdot hs_{i,r,t}^{\uparrow}) = \Pi_{r,t}^{\uparrow} \cdot hs_{i,r,t}^{\uparrow}$.

(20) is the weighted (by quantity) average of FO suppliers' own strike prices and the RT price.

Third, we estimate the RT credits/charges for FO buyers/sellers, respectively. A FO seller is charged the spread between the RT energy price and its own strike price:

$$\max\left(0, \pm \lambda_t^{RT} \mp C_{i,t}^{\uparrow|\downarrow}\right) \cdot \alpha_{r,t}^{\uparrow|\downarrow} \cdot hs_{i,r,t}^{\uparrow|\downarrow} \quad (21)$$

A FO buyer is then credited the spread between the RT energy price and the system-wide strike price:

$$\max\left(0, \pm \lambda_t^{RT} \mp VC_{i,t}^{\uparrow}\right) \cdot \overline{HD}_{i,r,t}^{\uparrow|\downarrow} \quad (22)$$

Note that in the equations above we have shown the process for calculating RT FO settlements by checking the trigger quantity condition for each FO buyer separately. Other variations could check the trigger quantity conditions using the net imbalance from all FO buyers.

## IV. NUMERICAL EXAMPLE

To highlight key characteristics of FOs, we define a simple test system and obtain results by solving a DA and multiple RT markets (one per scenario) for a single hour. We compare the results obtained to the outcome of a market design that includes an Imbalance Reserve product (IR), whose formulation is presented in Section IV-A. The test system is simple so that readers can easily replicate the results using models available in [33]. The system serves 200 MWh of load. It has one uncertain resource, with five equiprobable scenarios for its RT production, and five thermal generators (see Table I).

TABLE I
TEST SYSTEM: GENERATOR CHARACTERISTICS

| Thermal generators | ST1 | CT2 | CT3 | CT4 | CT5 |
|---|---|---|---|---|---|
| Capacity (MW) | 50 | 10 | 10 | 10 | 10 |
| Variable cost ($/MWh) | 20 | 35 | 50 | 60 | 70 |
| **Scenarios** | sc1 | sc2 | sc3 | sc4 | sc5 |
| RE (MW) | 131 | 141 | 155 | 165 | 172 |

The thermal fleet's ramping capability affects how much IR/FO each generator can offer. Using six distinct assumptions for the fleet's ramping capabilities helps us show how the relative performance of the two products (FO vs. IR) varies across systems (see Table II). Fleet1 is the most flexible, having no limits on its ability to change production. Fleets 2 to 6 have limited ramping capability, being able to ramp up or down to RT operating points within predefined neighborhoods of the DA energy schedules. The more expensive generators also have more ramping capability in fleets 2 and 3, while they have lower ramping capability in fleets 4-6.

TABLE II
RAMP RATE (MW/H)

| Thermal generators | ST1 | CT2 | CT3 | CT4 | CT5 |
|---|---|---|---|---|---|
| Fleet1 (MW/h) | 50 | 10 | 10 | 10 | 10 |
| Fleet2 (MW/h) | 6 | 8 | 10 | 10 | 10 |
| Fleet3 (MW/h) | 4 | 6 | 8 | 10 | 10 |
| Fleet4 (MW/h) | 10 | 10 | 10 | 8 | 6 |
| Fleet5 (MW/h) | 10 | 10 | 8 | 6 | 4 |
| Fleet6 (MW/h) | 20 | 1 | 1 | 0 | 0 |

### A. The Alternative: Imbalance Reserves

Many system operators already have or are in the process of adopting reserve products to manage DA-RT imbalances resulting from net load uncertainties. Hence, we simulate a simplified design for IR products that allows for an apples-to-apples comparison with FOs. The objective function minimizes the DA energy costs plus an estimate of the expected RT scarcity costs resulting from the chosen level $l$ of IR, the latter being captured by a downward demand curve for IR (scarcity cost $WTP_{l,t}$ varies per level $l$ in (23) and the steps of the demand curve are in the right-hand-side of (27),(28)). In addition to the DA energy balance constraint (24), which includes a virtual supplier ($VS_t$) at cost ($VBC$), constraints (25) and (26) procure enough upward and downward reserves ($RESREQ_t^{\uparrow,\downarrow}$) to manage imbalances for all scenarios included in the example. The sellers cannot provide more IR than what is technically possible given their rampability and maximum, and minimum capacities ((29)-(31)). RE can also sell IR by scheduling less DA generation than their deterministic forecast (which is then used as $P_{i,t}^{max}$).

$$\begin{aligned}\text{minimize} \quad & \sum_{i \in G, t \in T} C_{i,t} \cdot p_{i,t}^{DA} + \sum_{l \in L, t \in T} WTP_{l,t}^{\uparrow,\downarrow} \cdot short_{l,t}^{\uparrow,\downarrow} \\ & + \sum_t (D1 \cdot (d_t^{DA}) + D2 \cdot (d_t^{DA})^2) + VBC \cdot VS_t \\ \text{where} \quad & \Xi = \{\{p_{i,t}^{DA}, res_{i,t}^{\uparrow,\downarrow}, short_{l,t}^{\uparrow,\downarrow}, d_t^{DA}\} \geq 0, VS_t\}\end{aligned} \quad (23)$$

Subject to the following constraints:

$$\sum_{i \in G} p_{i,t}^{DA} + d_t^{DA} + VS_t = D_t \quad \forall t \in T \quad (24)$$

**Demand for imbalance reserves:** $\forall t \in T$

$$\sum_i res_{i,t}^{\uparrow} + \sum_l short_{l,t}^{\uparrow} \geq RESREQ_t^{\uparrow} \quad (25)$$

$$\sum_i res_{i,t}^{\downarrow} + \sum_l short_{l,t}^{\downarrow} \geq RESREQ_t^{\downarrow} \quad (26)$$

$$short_{l,t}^{\uparrow} \leq STEP_{l,t}^{\uparrow} \quad (27)$$

$$short_{l,t}^{\downarrow} \leq STEP_{l,t}^{\downarrow} \quad (28)$$

**For sellers of imbalance reserves:** $\forall i \in G, \forall t \in T$

$$res_{i,t}^{\uparrow,\downarrow} \leq u_{i,t} \cdot RR_i \quad (29)$$

$$p_{i,t}^{DA} + res_{i,t}^{\uparrow} \leq P_{i,t}^{max} \cdot u_{i,t} \quad (30)$$

$$p_{i,t}^{DA} - res_{i,t}^{\downarrow} \geq P_{i,t}^{min} \cdot u_{i,t} \quad (31)$$

## V. Results of FO and IR Comparison

We discuss the results in three subsections. In subsection V-A, we focus on the primal problem and we assess how effective the introduction of IR or FOs are in yielding an optimal DA schedule (i.e., expected cost-minimizing including penalties for unserved energy). In subsection V-B, we focus on the dual problem, and we assess how the introduction of IR or FOs affect price signals for flexibility in DA and energy prices in DA and RT. Last, in subsection V-C, we focus on settlements, and we assess how each of the two products (1) enables the system operator to be revenue-neutral and adhere to cost-causation principles; and (2) affects participants' flexibility-related payments and gross margins.

### A. Effectiveness of Day-ahead Schedules

Including FOs in the DA market is equally or more effective as inclusion of IRs, with the FO's probability-weighted RT system objective function being lower or equal to that of the IR problem (see column 1, Table III). In particular, FO and IR are equally cost-effective in two cases: when the system has either abundant flexibility (fleet 1) or a single DA operating point with adequate upward flexibility (fleet 6). In all other cases (fleets 2-5), IR is less cost-effective than FO because the IR formulation does not consider the probability-weighted cost increases or savings associated with deployment of flexibility up or down. In particular, IR's scheduling of ST1 at capacity (50 MW) for fleets 2-5 (see column ST1; table III) reveals IR's difficulty in valuing flexibility down by solely relying on scarcity pricing for down reserves. Similarly, IR's zero DA schedules for CT2 and CT3 for fleets 2-5 (see columns CT2/CT3; table III) illustrate IR's inability to anticipate some of the costs associated with deploying flexibility up.

TABLE III
PROBABILITY-WEIGHTED SYSTEM COST AND DA ENERGY SCHEDULES.

|  | System cost ($) | ST1 (MW) | CT2 (MW) | CT3 (MW) |
|---|---|---|---|---|
| fleet1 | 1,055/1,055 | 50/45 | 0/0 | 0/0 |
| fleet2 | 1,166/1,107 | 50/44 | 0/2 | 0/0 |
| fleet3 | 1,206/1,139 | 50/46 | 0/4 | 0/0.96 |
| fleet4 | 1,123/1,063 | 50/40 | 0/0.01 | 0/0 |
| fleet5 | 1,125/1,063 | 50/40 | 0/0 | 0/0.96 |
| fleet6 | 1,289/1,289 | 30.2/30.14 | 9/9 | 7.85/7.85 |

Table note: IR/FO results shown left and right of '/', respectively.

In contrast, the FO-based market finds that DA scheduling of CT3 is beneficial for fleets 3 and 5 because CT3 provides downward flexibility when its upward flexibility is not needed. For instance, for fleet 3, the scheduled amount of renewable energy (RE) is approximately 149 MW ($200-46-4-0.96$, Table III). Using the RE scenarios from Table I, we calculate the cumulative probability distribution of net load (NL) forecast error (see rows NL in table IV) and we find zero supply surplus with a cumulative probability of 80% (column:fleet 3, row 80% NL, Table IV). However, the demand for down FO with 80% probability is non-zero and equal to CT3's DA energy schedule (column:fleet 3, row 80% FO, Table IV). According to this result, CT3 would ramp down even in scenarios with RT NL higher than DA NL (e.g., sc2)! To make up for CT3's ramping down, another unit has to ramp up in sc2. That's why the demand for up FO at a 40% cumulative probability is higher than the corresponding NL forecast error.

In simple terms, the FO market recognizes that under certain scenarios (far from the extremes), it is cost effective to deploy flexibility in both directions (up from ST1 and down from CT3). This result shows that the economic value of (and demand for) flexibility depends not only on net load forecast error but also on the pool of reserve suppliers. Considering the pool of reserve suppliers during the construction of demand curves for reserve products such as IRs would be impractically complicated, and perhaps impossible. However, it is relatively easy for the FO model because the latter product endogenously yields the demand for flexibility considering the distribution of trigger quantities (as submitted by FO buyers) and the balancing costs reflected in the RT strike prices of FO suppliers.

TABLE IV
CUMULATIVE FO DEMAND AND NL FORECAST ERROR (MW)

| Probability | fleet 1 | fleet 2 | fleet 3 | fleet 4 | fleet 5 | fleet 6 |
|---|---|---|---|---|---|---|
| 20% FO | 24/17 | 23/8 | 18/9 | 29/10 | 28/11 | 22/19 |
| 20% NL | 24/17 | 23/18 | 18/23 | 29/12 | 28/13 | 22/19 |
| 40% FO | 14/10 | 13/8 | 9/9 | 19/5 | 19/6 | 14/12 |
| 40% NL | 14/10 | 13/11 | 8/16 | 19/5 | 18/6 | 12/12 |
| 60% FO | 0/0 | 1/2 | 0/6 | 5/0 | 5/1 | 0/2 |
| 60% NL | 0/0 | 0/1 | 0/6 | 5/0 | 4/0 | 0/2 |
| 80% FO | 0/0 | 0/0 | 0/1 | 0/0 | 0/1 | 0/2 |
| 80% NL | 0/0 | 0/0 | 0/0 | 0/0 | 0/0 | 0/0 |

Table note: Upward/downward shown left and right of '/', respectively.

### B. Price Signals

Including flexibility products (IR or FO) in the DA market will provide price signals for flexibility. As shown in Table V, the two price signals are quite different. In case of upward flexibility, both signals have two levels in the six fleets studied. However, those two levels are different because IR prices flexibility considering only the opportunity cost with respect to the DA energy schedule, while FO also considers the price at which flexibility will be provided (i.e., strike price for FO suppliers). The zero price for IR up flexibility in fleets 1-5 does not indicate that there are no profitable opportunities for upward flexibility in RT. It just indicates that enough upward flexibility exists in the system without adjusting the least-DA-cost energy schedule. In contrast, the upward flexibility price for tier 2 in FO indicates the potential value that a flexible resource could provide to the market in RT. In case of downward flexibility, the prices from FO are much more dynamic. The IR down prices are 0 for all fleets, whereas the FO down prices for tier 2 have three levels. The down price is negative because it reflects the probability-weighted price a flexible resource would have to pay in RT to outsource its energy schedules to another market participant. The FO prices provide a clear signal about the value of upward and downward flexibility, whereas the IR prices provide a harder to decipher signal through the prices for IR up and DA energy.

TABLE V
PRICE SIGNALS FOR FLEXIBILITY ($/MW)

|  | fleet1 | fleet2 | fleet3 | fleet4 | fleet5 | fleet6 |
|---|---|---|---|---|---|---|
| Up | 0/17 | 0/17 | 0/17 | 0/17 | 0/17 | 30/38 |
| Down | 0/-12 | 0/-4 | 0/-4 | 0/-8 | 0/-8 | 0/-12 |

Table notes: 1) IR/FO results shown left and right of '/', respectively. 2) The FO has a differentiated price for flexibility per tier. We here show the prices for tier 2 in both up and down direction.

Turning to energy prices, we note that whereas DA and RT energy prices converge on average in simulations with either product (IR or FO), the level at which they converge is different. As shown in Table VI, the average energy prices in FO simulations can be higher, lower, or equal to those in IR simulations. Overall, IR simulations result in higher or lower average energy prices, respectively, when they rely more on expensive and inexpensive resources for upward and downward flexibility. For example, IR simulations for fleets 2 and 3 yield higher energy prices because they turn on more expensive units in sc1 and sc2. Similarly, IR simulations for fleets 4 and 5 yield lower energy prices because IR simulations curtail more MWh from the inexpensive RE in sc4.

TABLE VI
ENERGY PRICES: DA AND SCENARIO-SPECIFIC RT ($/MWH)

|         | DA    | sc1     | sc2   | sc3   | sc4   | sc5   |
|---------|-------|---------|-------|-------|-------|-------|
| fleet 1 | 29/29 | 50/50   | 35/35 | 20/20 | 20/20 | 20/20 |
| fleet 2 | 26/21 | 60/50   | 50/35 | 20/20 | 0/0   | 0/0   |
| fleet 3 | 22/21 | 60/50   | 50/35 | 0/20  | 0/0   | 0/0   |
| fleet 4 | 21/25 | 50/50   | 35/35 | 20/20 | 0/20  | 0/0   |
| fleet 5 | 23/25 | 60/50   | 35/35 | 20/20 | 0/20  | 0/0   |
| fleet 6 | 50/50 | 170/170 | 20/20 | 20/20 | 20/20 | 20/20 |

Table note: IR/FO results shown left and right of '/', respectively.

The convergence between DA and RT energy prices is also achieved in different ways. In simulations with IR, the convergence is possible only by including a virtual bidder (VB), who participates in DA and has perfect foresight of the distribution of RT system conditions.[3] Inclusion of the VB is necessary for convergence because the IR design does not have any information on the RT costs of the flexible resources. On the contrary, the FO simulations rely on strike prices to endogenously consider the uncertain resource's DA-RT arbitrage problem through equations (7) and (4b)-(4d), resulting in energy prices in the DA and RT markets moving towards convergence even without including a VB. As shown in Table VII, the uncertain renewable resource is always scheduled at its expected value (the maximum we allow) in DA IR. In contrast, the uncertain renewable resource adjusts its schedule depending on the supply of flexibility in FO. For example, RE turns out to be scheduled below and above the median in fleets 2&3 and 4&5 when upward and downward flexibility is relatively more expensive to buy, respectively. For a more general discussion on price convergence, please refer to Appendix B, which builds upon the DA and RT problems, presented in Section III and Appendix A, respectively.

TABLE VII
RE AND VIRTUAL DA ENERGY SUPPLY (MW)

|         | fleet1     | fleet2     | fleet3     |
|---------|------------|------------|------------|
| RE      | 152.8/155  | 152.8/154  | 152.8/149  |
| Virtual | -2.8/0     | -2.8/0     | -2.8/0     |

|         | fleet4     | fleet5     | fleet6     |
|---------|------------|------------|------------|
| RE      | 152.8/160  | 152.8/159  | 152.8/153  |
| Virtual | -2.8/0     | -2.8/0     | +0.11/0    |

Table note: IR/FO results shown left and right of '/', respectively.

It is important to note that subsection V-A showed that the expected system cost is no lower in the IR cases, and is often higher. Thus, convergence of DA and RT prices is not sufficient to guarantee cost-efficiency.

---

[3]We determine the price and amount of virtual bidding by solving a mixed complementarity problem.

## C. Settlements

*1) Revenue adequacy:* By design, the FO settlement scheme ensures that the system operator's cash outlay in FO settlements equals its receipts, making it revenue-neutral. On the other hand, the IR settlement scheme does not guarantee revenue neutrality. The results for fleet 6 clearly show the difference between the two products. In the FO market, FO buyers (here RE) pay FO sellers in DA, and FO buyers receive option payoffs in RT. Hence, the system operator is always revenue-adequate (see the last row of Table VIII). However, in the case of IR, the system operator pays flexible resources in DA ($654 here) and charges the resources with RT-DA imbalances in RT. The ISO pays flexible resources $30/MW when it reserves capacity to manage the worst-case upward imbalance (21.8 MW). This worst-case upward imbalance occurs only in one RT scenario (sc1) and the system operator then charges the RE $654, which is equal to the amount the ISO paid the flexible resources DA. However, in the other four RT scenarios (sc2-sc5), the ISO charges the RE for lower than the worst-case upward imbalance, yielding a lower expected revenue from RE (131+71) than the credits to the flexible resources (654).

TABLE VIII
FLEET 6: IR/FO-RELATED PROBABILITY-WEIGHTED SETTLEMENTS IN $

|     | DA      | RT sc1   | RT sc2 | RT sc3 | RT sc4 | RT sc5 |
|-----|---------|----------|--------|--------|--------|--------|
| ST1 | 594/596 | 0/-596   | 0/0    | 0/0    | 0/0    | 0/0    |
| CT2 | 30/39   | 0/-27    | 0/-3   | 0/-3   | 0/-3   | 0/-3   |
| CT3 | 30/48   | 0/-24    | 0/-6   | 0/-6   | 0/-6   | 0/-6   |
| RE  | 0/-683  | -131/647 | -71/9  | 0/9    | 0/9    | 0/9    |
| ISO | -654/0  | 131/0    | 71/0   | 0/0    | 0/0    | 0/0    |

Results from Table VIII illustrate another key difference between the two products. The probability-weighted sum of FO-related charges and credits is approximately zero per participant, meaning that each participant's benefits are equal on average to their costs from buying or selling FOs. The intuition behind this result is that the marginal FO price at each tier will be equal to the expected value of RT energy prices for all scenarios with the FO exercised. Hence, each FO seller receives for each MW in a tier precisely the margin it foregoes in RT, assuming that the discrete distribution is 100% correct. For a mathematical analysis of this result, please refer to Appendix C.

On the contrary, the IR scheme for systems like fleet 6 tends to over-compensate flexible resources for their opportunity costs in DA because it overlooks the ability of flexible resources to realize profits in RT. For example, in fleet 6, the IR scheme pays ST1 $594 (30·21.8) to secure 21.8 MW of upward flexibility because ST1 foregoes $30/MWh (50-20) of DA energy profits when it sells IR. However, the IR scheme fails to account for the fact that ST1 will also make a gross margin in sc1 equal to $30/MWh (0.2·(170-20)). As a result, the IR scheme overestimates ST1's opportunity costs. Interestingly, the FO scheme estimates similar DA opportunity costs for ST1. However, the FO scheme does not overpay ST1 because ST1 gives back to FO buyers any excess of RT strike price gross margins in RT (see sc1, TableVIII).

*2) Flexibility-related Payments and Gross Margins:* The intermittency of flexibility-related revenues also differs for

the two products. Overall, the rewards for flexible resources tend to be more intermittent for supplying IR vs. providing FO (see Table IX). For instance, in the FO simulations, CT2 consistently makes positive margins in all five scenarios for fleets 1-5, whereas in the IR simulations, it only yields positive margins in one to two scenarios. For fleet 6, both products proactively reward flexibility in DA in every scenario due to the presence of lost opportunity costs related to DA energy sales. All FO suppliers with positive DA energy schedules have non-negative gross margins (i.e., they recover their costs). For a more general (mathematically-based) discussion on cost recovery, please refer to Appendix C.

TABLE IX
NUMBER OF SCENARIOS WITH POSITIVE GROSS MARGINS

|     | fleet1 | fleet2 | fleet3 | fleet4 | fleet5 | fleet6 |
| --- | --- | --- | --- | --- | --- | --- |
| ST1 | 5/5 | 5/5 | 5/5 | 5/5 | 5/5 | 5/5 |
| CT2 | 1/5 | 2/5 | 2/5 | 1/5 | 1/5 | 5/5 |
| CT3 | 0/0 | 1/0 | 1/5 | 0/0 | 1/5 | 5/5 |

Table note: IR/FO results shown left and right of '/', respectively.

For resources with imbalances (here RE), we observe that the FO RT payoffs help reduce the the range of RE gross margins (relative to the mean gross margin) across scenarios, thereby addressing RE's hedging needs (see Table X).

TABLE X
RE GROSS MARGIN PER SCENARIO (NORMALIZED BY THE MEAN OVER ALL SCENARIOS GROSS MARGIN)

|     | fleet1 | fleet2 | fleet3 |
| --- | --- | --- | --- |
| sc1 | 0.78/0.83 | 0.74/0.76 | 0.69/0.77 |
| sc2 | 0.94/0.94 | 0.94/0.93 | 0.93/0.94 |
| sc3 | 1.05/1.04 | 1.12/1.07 | 1.13/1.08 |
| sc4 | 1.1/1.08 | 1.1/1.12 | 1.13/1.1 |
| sc5 | 1.13/1.12 | 1.1/1.12 | 1.13/1.1 |
|     | fleet4 | fleet5 | fleet6 |
| sc1 | 0.73/0.79 | 0.69/0.79 | 0.48/0.92 |
| sc2 | 0.96/0.93 | 0.98/0.93 | 1.04/0.97 |
| sc3 | 1.12/1.05 | 1.12/1.05 | 1.13/1.01 |
| sc4 | 1.1/1.1 | 1.11/1.1 | 1.16/1.04 |
| sc5 | 1.1/1.13 | 1.11/1.13 | 1.18/1.06 |

Table note: IR/FO results shown left and right of '/', respectively.

Finally, we analyze how different the probability-weighted gross margins are for all participants (flexible resources, RE, consumers) between the two designs (see Table XI). Whereas the FO design has lower or equal costs to IR in all fleets, the gross margins for participants do not change in a consistent direction among fleets. This is true for flexible resources, RE, and consumers. For example, flexible resources receive the same probability-weighted gross margin in fleet 1 with either product. However, for fleets 2 and 3, the probability-weighted gross margin is lower under FO compared to IR for all flexible resources because the energy prices converge to a much lower level, benefiting consumers. For fleets 4 and 5, inexpensive flexible resources (ST1) are better off with FO while expensive flexible resources (CT2, CT3) are worse off with FO. Last, in fleet 6, all flexible resources yield higher probability-weighted gross margins in IR. This difference in gross margins is mainly funded by the ISO and its revenue-inadequate reserve settlement scheme.

In conclusion, flexible resources access a less intermittent revenue stream in FO. However, their incentives, in terms of total gross margins, to offer FOs vary. For fleets 4 and 5, it is clear that ST1 has an incentive to offer FOs. However, for fleets 2, 3, and 6, flexible resources do not have a clear incentive to offer FOs as the energy price levels are lower under FO or they receive additional revenues through the IR revenue-inadequate settlement scheme. In those cases, strategic flexible resources might prefer to leave the FO market, and a regulator might need to consider introducing rules on behalf of consumers that would help achieve the outcome of a perfectly competitive market with truthful bids.

TABLE XI
PROBABILITY-WEIGHTED GROSS MARGINS OF PARTICIPANTS ($)

|     | fleet1 | fleet2 | fleet3 |
| --- | --- | --- | --- |
| ST1 | 450/450 | 348/146 | 148/115 |
| CT2 | 30/30 | 64/30 | 48/30 |
| CT3 | 0/0 | 20/0 | 16/0.25 |
| RE | 4265/4265 | 3602/2919 | 2982/2919 |
| LOAD | -5800/-5800 | -5200/-4202 | -4400/-4204 |
| ISO | 0/0 | 0/0 | 0/0 |
|     | fleet4 | fleet5 | fleet6 |
| ST1 | 130/330 | 230/332 | 2094/1500 |
| CT2 | 30/30 | 50/30 | 204/174 |
| CT3 | 0/0 | 16/0.25 | 78/48 |
| RE | 2917/3579 | 3179/3583 | 6784/6986 |
| LOAD | -4200/-5002 | -4600/-5008 | -10000/-10000 |
| ISO | 0/0 | 0/0 | -452/0 |

Table note: IR/FO results shown left and right of '/', respectively.

## VI. CONCLUSIONS

We introduce a novel product, Flexibility Options, for managing DA to RT imbalances in ISO-operated DA markets. Co-optimizing the procurement of this product with DA energy is crucial as the demand for and availability of Flexibility Options depends on DA energy schedules.

Whereas this design is currently proposed for managing DA-to-RT imbalances, it could generally be beneficial for managing risks related to imbalances with at least four characteristics. First, for the product to be directly applicable, the imbalances should be settled in a centralized manner with pay-as-clear (uniform) pricing. In other words, the product is not relevant for imbalances that are managed via ad-hoc processes triggered by events (e.g., sudden loss of an online unit). Second, relatively sharp and well-calibrated probabilistic forecasts for the distribution of imbalances are likely to result in more accurate pricing when a product like the one proposed here is used. Third, implementation of such a product will likely be most useful when the management of imbalances can result in significant probability-weighted activation costs of reserved flexibility (i.e., balancing costs). Last, a product like FO will likely give more benefits when there are multiple resources that could manage imbalances, and activation costs among suppliers of flexibility vary.

For management of DA-to-RT imbalances, we find that Flexibility Options result in equally or more cost-effective day-ahead schedules than traditional reserve products. Flexibility Options might also be preferred to traditional reserve products because they can price flexibility more accurately by satisfying the hedging needs of uncertain resources while accounting for balancing costs of flexible resources. This has been demonstrated in cases where imbalances follow a discrete probability distribution and are perfectly correlated among FO buyers. The revenue adequacy of its settlement scheme is another desirable property of Flexibility Options, which gives the product a competitive edge compared to traditional reserve products whose costs have to be socialized in part or totally.

The introduction of the product offers benefits to all parties involved. For FO sellers, the product provides a revenue stream that values flexibility, is less intermittent than RT balancing margins, and guarantees cost recovery per scenario. For FO buyers, the product provides hedging and ensures cost recovery in expectation. For system operators, the product facilitates procurement of flexibility in a revenue-neutral manner.

However, these benefits come with responsibilities. The FO sellers must commit to financially binding strike prices. The FO buyers must project the probability distribution of their RT outputs. The system operator must facilitate FO trading. The extent to which the benefits outweigh the burden of responsibilities will be informed by stakeholder discussions, formal cost-benefit analysis studies similar to ones conducted in the past for market enhancements of similar scope [34], and eventually experience.

In follow-up work, we plan to report results that quantify the benefits of Flexibility Options for large-scale systems. For large-scale systems, the main challenge is to strike a balance between additional complexity and added value. Several choices will have to be made to account for this trade-off between complexity and value. Two such choices are the number of discrete levels for the distribution of imbalances and the method by which the system operator aggregates the demand for Flexibility Options. Another aspect that is important for large-scale systems is transmission congestion. Congestion can challenge the physical delivery of Flexibility Options or can influence the value of the hedges, as the presented formulation is only useful for hedging the energy portion of the locational marginal price. Last but not least, the model presented here is for a single period and large-scale application will need to account for inter-temporal ramping constraints to ensure physical delivery of Flexibility Options.

Another interesting area for future research on this product relates to the participation of financial entities as both FO buyers and sellers. Similar to virtual bidders in US markets, participation of financial entities in FO trading could enhance the benefits of trading in the presence of heterogeneous risk attitudes and market power.

## APPENDIX A: RT PROBLEM

The problem formulation of III-A co-optimizes the DA market with probability-weighted scenario-specific RT markets. Here, we provide the formulation for a RT market scenario.

$$\begin{aligned} \text{minimize} \quad & Objective_s^{RT} \\ \text{where} \quad & \Xi = \{d_{s,t}^{RT}, p_{i,s,t}^{RT,+}, sd_{i,s,t}^{RT,+}, p_{i,s,t}^{RT,-}, sd_{i,s,t}^{RT,-}\} \end{aligned} \quad (A.1)$$

The $Objective_s^{RT}$ is defined as follows:

$$\sum_{i \in G^S, t \in T} \left( C_{i,t}^{\uparrow} \cdot p_{i,s,t}^{RT,+} - C_{i,t}^{\downarrow} \cdot p_{i,s,t}^{RT,-} \right) \quad (A.2a)$$

$$+ \sum_{i \in G^B, t \in T} \left( C_{i,t} \cdot (p_{i,s,t}^{RT,+} - p_{i,s,t}^{RT,-}) \right) \quad (A.2b)$$

$$+ \sum_{i \in G^B, t \in T} \left( (\pm \widetilde{VC_{i,t}^{\uparrow|\downarrow}} \mp C_{i,t}) \cdot sd_{i,s,t}^{RT,-|+} \right) \quad (A.2c)$$

$$+ (D1 \cdot (d^{DA} + d_s^{RT}) + D2 \cdot (d^{DA} + d_s^{RT})^2) \quad (A.2d)$$

$$- (D1 \cdot (d^{DA}) + D2 \cdot (d^{DA})^2) \quad (A.2e)$$

The optimization includes the following constraints:
**Balancing constraint**

$$\sum_{i \in G} p_{i,s,t}^{RT,+} + d_{s,t}^{RT} = \sum_{i \in G} p_{i,s,t}^{RT,-} \quad \forall t \in \Omega_T \quad (\lambda_t^{RT}) \quad (A.3)$$

**Uncertain participants:** $\forall i \in G^B, \forall t \in T$

$$p_{i,s,t}^{RT,+} - p_{i,s,t}^{RT,-} + sd_{i,s,t}^{RT,+} - sd_{i,s,t}^{RT,-} = P_{i,s,t} - p_{i,t}^{DA} \quad (A.4)$$

**Flexible participants:** $\forall i \in G^S, \forall t \in T$

$$-RR_{i,t} \cdot u_{i,t} \leq -p_{i,s,t}^{RT,-} \quad p_{i,s,t}^{RT,+} \leq RR_{i,t} \cdot u_{i,t} \quad (A.5)$$

$$p_{i,t}^{DA} + p_{i,s,t}^{RT,+} \leq P_{i,t}^{max} \cdot u_{i,t} \quad (A.6)$$

$$p_{i,t}^{DA} - p_{i,s,t}^{RT,-} \geq P_{i,t}^{min} \cdot u_{i,t} \quad (A.7)$$

## APPENDIX B: CONVERGENCE OF DA AND RT PRICES

We study the convergence between DA and RT energy prices to understand if the introduction of FOs in DA creates asymmetries between the two markets.

For the test system of Section IV, we study DA to RT energy price convergence under 1,000 random sets of values for ramp rates, strike up and strike down prices. The values are drawn from uniform distributions between: 0 and the plant's capacity for ramp rates; the DA variable cost and double that for strike-up prices, zero and the DA variable cost for strike-down prices.

For all tests, DA energy prices are approximately equal to the average RT energy prices. Moreover, we observe that for the pair of DA and RT solutions, the primal decision variables of DA can be combined to yield a RT optimal solution and the dual decision variables of RT can be combined to yield a DA optimal dual as shown in B.1-B.10.

Primal decision variables:
For each FO seller $i \in G^S$:

$$p_{i,s,t}^{RT*,+} = \sum_{r \geq s} hs_{i,r,t}^{\uparrow} \quad (B.1)$$

$$p_{i,s,t}^{RT*,-} = \sum_{r \leq s-1} hs_{i,r,t}^{\downarrow} \quad (B.2)$$

For each FO buyer $i \in G^B$:

$$p_{i,s,t}^{RT*,-} - p_{i,s,t}^{RT*,+} = \sum_{r \geq s} hd_{i,r,t}^{\uparrow} - \sum_{r \leq s-1} hd_{i,r,t}^{\downarrow} \quad (B.3)$$

$$sd_{i,s,t}^{RT*,-} - sd_{i,s,t}^{RT*,+} = \sum_{r \geq s} sd_{i,r,t}^{\uparrow} - \sum_{r \leq s-1} sd_{i,r,t}^{\downarrow} \quad (B.4)$$

Dual decision variables. For RT duals, superscript is the number of associated constraint:

$$\lambda_{r,t}^{\uparrow} \approx \sum_{s \leq r} \Pi_s \cdot \lambda_{s,t}^{RT,*} \quad \forall r | \sum_i hd_{i,r,t}^{\uparrow} > 0 \quad (B.5)$$

$$\lambda_{r,t}^{\downarrow} \approx -\sum_{r+1 \leq s} \Pi_s \cdot \lambda_{s,t}^{RT,*} \quad \forall r | \sum_i hd_{i,r,t}^{\downarrow} > 0 \quad (B.6)$$

$$\lambda_{i,t}^{11} = max(0, \lambda_t^{DA} - C_{i,t}) \quad (B.7)$$

$$\lambda_{i,t}^{12} = max(0, \quad C_{i,t} - \lambda_t^{DA}) \quad (B.8)$$

$$\lambda_{i,t}^{10R} \approx max(0, \sum_{\substack{k \in \{5R,6\}, \\ s | \sum_{r \geq s} hs_{i,r,t}^{\uparrow} \geq 0}} \Pi_s \cdot \lambda_{i,s,t}^{k,RT} - \lambda_{i,t}^{11}) \quad (B.9)$$

$$\lambda_{i,t}^{10L} \approx max(0, \sum_{\substack{k \in \{5L,7\}, \\ s | \sum_{r \leq s-1} hs_{i,r,t}^{\downarrow} \geq 0}} \Pi_s \cdot \lambda_{i,s,t}^{k,RT} - \lambda_{i,t}^{12}) \quad (B.10)$$

## APPENDIX C: COST RECOVERY

Here we discuss how FO suppliers and buyers recover their costs in each scenario and in expectation, respectively. The gross margin in each RT scenario $s$ for a supplier of FOs consists of four terms. The first two terms are from DA results: (C.1a) DA energy margin, (C.1b) DA option premiums. Terms (C.1c) and (C.1d) are based on RT results and occur when $i$ deploys flexibility up and down, respectively. Each of these latter components consists of two parts: the gross margin from participating in the RT energy market and the FO payoff that the entity has to remit to the system operator.

$$(\lambda_t^{DA} - C_{i,t}) \cdot p_{i,t}^{DA} \quad \text{(C.1a)}$$

$$+ \sum_r (\lambda_{r,t}^\uparrow - \Pi_{r,t}^\uparrow \cdot C_{i,t}^\uparrow) \cdot hs_{r,t}^\uparrow + (\lambda_{r,t}^\downarrow + \Pi_{r,t}^\downarrow \cdot C_{i,t}^\downarrow) \cdot hs_{r,t}^\downarrow \quad \text{(C.1b)}$$

$$+ (\lambda_{s,t}^{RT} - C_{i,t}^\uparrow) \cdot p_{i,s,t}^{RT,+} - \sum_{s \leq r} \alpha_{r,t}^\uparrow \cdot \max(0, \lambda_{s,t}^{RT} - C_{i,t}^\uparrow) \cdot hs_{r,t}^\uparrow \quad \text{(C.1c)}$$

$$- (\lambda_{s,t}^{RT} - C_{i,t}^\downarrow) \cdot p_{i,s,t}^{RT,-} - \sum_{r+1 \leq s} \alpha_{r,t}^\downarrow \cdot \max(0, C_{i,t}^\downarrow - \lambda_{s,t}^{RT}) \cdot hs_{r,t}^\downarrow \quad \text{(C.1d)}$$

The supplier recovers its costs because all terms of (C.1) are non-negative. Term (C.1.a) is non-negative unless $i$ generates $P_{min}$. In that case, $\lambda_{i,t}^{12}$ is positive and uplift payments support $i$. Term (C.1.b) is non-negative because of the complementarity conditions wrt $hs^{\uparrow,\downarrow}$. Similarly, the first terms of (C.1.c) and (C.1.d) are non-negative because of the complementarity conditions wrt $p_{i,s,t}^{RT,+|-}$. The second terms of (C.1.c) and (C.1.d) should not be higher than the first terms if (B.1) and (B.2) hold. Using (B.5) and (B.6), for $\alpha_{r,t} = 1$ we show step by step below how in expectation (consider all RT scenarios with their probabilities), the $2^{nd}$ terms of (C.1c) and (C.1d) cancel term (C.1b), i.e., the FO premiums (C.1b) converge to the RT FO payoffs.

Starting from (C.1b), we replace the dual variables of the FO supply-demand balance at each tier with the right hand side from (B.5) and (B.6).

*1st term from C1.b:* $\sum_r (\lambda_{r,t}^\uparrow - \Pi_{r,t}^\uparrow \cdot C_{i,t}^\uparrow) \cdot hs_{r,t}^\uparrow$

We first replace the dual variables of FO up supply-demand balance at each tier with the right hand side from (B.5).
$\sum_r ((\sum_{s \leq r} \Pi_s \cdot \lambda_{s,t}^{RT}) - \Pi_{r,t}^\uparrow \cdot C_{i,t}^\uparrow) \cdot hs_{r,t}^\uparrow$.

Next, we replace the probability of exercising an up FO with the cumulative probability of the scenarios whose upper operating limit is lower than the trigger quantity of the tier.
$\sum_r (\sum_{s \leq r} \Pi_s \cdot \lambda_{s,t}^{RT} - \sum_{s \leq r} \Pi_s \cdot C_{i,t}^\uparrow) \cdot hs_{r,t}^\uparrow$

Rearranging the tiers with scenarios, we get:
$\sum_s \Pi_s \cdot (\lambda_{s,t}^{RT} - C_{i,t}^\uparrow) \cdot (\sum_{s \leq r} hs_{r,t}^\uparrow)$ This expression is identical to the 2nd term of (C.1c)) given that the up FOs are activated only when $\lambda_{s,t}^{RT} \geq C_{i,t}^\uparrow$.

*2nd term from C1.b:* $\sum_r (\lambda_{r,t}^\downarrow + \Pi_{r,t}^\downarrow \cdot C_{i,t}^\downarrow) \cdot hs_{r,t}^\downarrow$ We first replace the dual variables of FO up supply-demand balance at each tier with the right hand side from (B.6).
$\sum_r (-\sum_{r \leq s-1} \Pi_s \cdot \lambda_{s,t}^{RT} + \Pi_{r,t}^\downarrow \cdot C_{i,t}^\downarrow) \cdot hs_{r,t}^\downarrow$.

Next, we replace the probability of exercising a down FO with the cumulative probability of the scenarios whose upper operating limit is higher than the trigger quantity of the tier.
$\sum_r (-\sum_{r \leq s-1} \Pi_s \cdot \lambda_{s,t}^{RT} + \sum_{r \leq s-1} \Pi_s \cdot C_{i,t}^\downarrow) \cdot hs_{r,t}^\downarrow$

Rearranging the tiers with scenarios, we get:
$\sum_s \Pi_s \cdot (C_{i,t}^\downarrow - \lambda_{s,t}^{RT}) \cdot (\sum_{r \leq s+1} hs_{r,t}^\downarrow)$ This expression is identical to the 2nd term of (C.1d)) given that there is no simultaneous activation of FOs from the same supplier $i$, and that the down FOs are activated only when $\lambda_{s,t}^{RT} \leq C_{i,t}^\downarrow$.

Note that the remaining terms (C.1a and 1st terms of C.1c and C.1d) are equal to the probability-weighted gross margin $i$ would make if it did not provide FOs and the prices remain unchanged.

For a FO buyer, the gross margin per scenario is as follows.

$$(\lambda_t^{DA} - C_{i,t}) \cdot p_{i,t}^{DA} \quad \text{(C.2a)}$$

$$\mp \sum_r \frac{\sum_{i \in G^S} \left(\lambda_{r,t}^{\uparrow|\downarrow} \mp \Pi_{r,t}^{\uparrow|\downarrow} \cdot C_{i,t}^{\uparrow|\downarrow}\right) \cdot hs_{i,r,t}^{\uparrow|\downarrow}}{\sum_{i \in G^B} hd_{i,r,t}^{\uparrow|\downarrow}} \cdot hd_{r,t}^{\uparrow|\downarrow} \quad \text{(C.2b)}$$

$$- (\lambda_t^{RT} - C_{i,t}) \cdot p_{i,s,t}^{RT,-} + \sum_{r \leq s} (\lambda_t^{RT} - VC_{r,t}^\uparrow) \cdot hd_{i,r,t}^\uparrow \quad \text{(C.2c)}$$

$$+ (\lambda_t^{RT} - C_{i,t}) \cdot p_{i,s,t}^{RT,+} + \sum_{r \geq s+1} (VC_{r,t}^\downarrow - \lambda_t^{RT}) \cdot hd_{i,r,t}^\downarrow \quad \text{(C.2d)}$$

Using (B.4) and (B.5), for $\alpha_{r,t} = 1$, it is easy to see that in expectation the $2^{nd}$ terms of (C.2c) and (C.2d) cancel term (C.2b). In that case, the remaining terms are equal to the probability-weighted gross margin $\widetilde{\text{unit } i}$ would make if it did not buy FOs and it is positive if $\widetilde{VC_{i,t}^\downarrow - C_{i,t}}$ is positive.

## APPENDIX D: THE ROLE OF M

For each FO buyer, the formulation has $2 * R$ decision variables for FOs and $|R|+1$ constraints limiting the volume of FOs of different tiers that can be exercised in each scenario. Therefore, unless there is some differentiation of FO strike prices at different tiers, alternative optimal solutions exist.

For instance, the FO baskets in Table XII are alternative optima for FO seller ST1 in fleet 6 as its DA schedule is 30.14 MWh and 31.46 MWh, when M is 0.01 and 0, respectively. Both baskets yield the same RT schedules for all units including ST1 (50, 40.14, 30.14, 20.14, 13.14) in sc1-sc5. The absolute volume of ST1's FOs exercised in any scenario is higher when M=0. The high FO volume unnecessarily complicates the problem, potentially introducing transaction costs. To choose among alternative optima the solution with the lowest absolute FO volume activated per scenario, we added term (4e) in the objective function.

TABLE XII
FLEET6: ALTERNATIVE OPTIMA FOR ST1 FO SALES (IN MW)

|       | M=0.01 ($hs^\uparrow$) | M=0.01 ($hs^\downarrow$) | M=0 ($hs^\uparrow$) | M=0 ($hs^\downarrow$) |
|-------|------------------------|--------------------------|---------------------|-----------------------|
| tier1 | 5.86                   | 0                        | 3.18                | 2.68                  |
| tier2 | 14                     | 0                        | 5.89                | 8.11                  |
| tier3 | 0                      | 10                       | 5.05                | 4.95                  |
| tier4 | 0                      | 7                        | 4.42                | 2.58                  |

## ACKNOWLEDGMENT

This work was authored in part by the National Renewable Energy Laboratory, operated by Alliance for Sustainable Energy, LLC, for the U.S. Department of Energy (DOE) under Contract No. DE-AC36-08GO28308. Funding provided by the Advanced Research Projects Agency-Energy (ARPA-e) under the Performance-based Energy Resource Feedback, Optimization, and Risk Management (PERFORM) program and a Leverhulme International Professorship. The views expressed in the article do not necessarily represent the views of the DOE or the U.S. Government. The U.S. Government retains and the publisher, by accepting the article for publication, acknowledges that the U.S. Government retains a nonexclusive, paid-up, irrevocable, worldwide license to publish or reproduce the published form of this work, or allow others to do so, for U.S. Government purposes. Specific statements may not necessarily represent the views of any one authors' employers.

The authors gratefully acknowledge Mohamed AlAshery and Jairo Cervantes for their contributions early in the project. The authors also gratefully acknowledge ARPA-E staff Joseph King, Ashley Arigoni, Richard O' Neill, Richard Wilson, and Jonathan Glass for their guidance and support. The authors thank Yingchen 'YC' Zhang, Erik Ela, Yonghong Chen, participants in EPRI's Price Formation Working Group and other ARPA-E PERFORM teams for their feedback on earlier versions of this work.

## References

[1] G. Bautista-Alderete and K. Zhao, "Day-ahead market enhancements analysis," California Independent System Operator, 2022. [Online]. Available: http://www.caiso.com/InitiativeDocuments/Day-AheadMarketEnhancementsAnalysisReport-Jan24-2022.pdf

[2] F. Billimoria, "Over the edge—energy risk trading in a negative demand environment," *IAEE Energy Forum*, vol. 2, 2021.

[3] S. Pineda and A. Conejo, "Managing the financial risks of electricity producers using options," *Energy Econ.*, vol. 34, no. 6, pp. 2216–2227, 2012.

[4] K. Alshehri, S. Bose, and T. Başar, "Cash-settled options for wholesale electricity markets," *IFAC-PapersOnLine*, vol. 50, no. 1, pp. 13 605–13 611, 2017, 20th IFAC World Congress.

[5] J. Kazempour and B. F. Hobbs, "Value of flexible resources, virtual bidding, and self-scheduling in two-settlement electricity markets with wind generation — Part I: Principles and competitive model," *IEEE Trans. Power Syst.*, vol. 33, no. 1, pp. 749–759, 2018.

[6] J. M. Morales, M. Zugno, *et al.*, "Electricity market clearing with improved scheduling of stochastic production," *Eur. J. Oper. Res.*, vol. 235, no. 3, pp. 765–774, 2014.

[7] J. Bushnell, S. M. Harvey, and B.F. Hobbs, "Opinion on Energy Imbalance Market (EIM) resource sufficiency evaluation enhancements, Phase 2," MSC of the California ISO, Revised Draft, October 23, 2022.

[8] V. Dvorkin, S. Delikaraoglou, and J. M. Morales, "Setting reserve requirements to approximate the efficiency of the stochastic dispatch," *IEEE Trans. Power Syst.*, vol. 34, no. 2, pp. 1524–1536, 2019.

[9] A. Papavasiliou and S. Oren, "Multiarea stochastic unit commitment for high wind penetration in a transmission constrained network," *Oper. Res.*, vol. 61, no. 3, pp. 578–592, 2013.

[10] L. A. Roald, D. Pozo, *et al.*, Power systems optimization under uncertainty: A review of methods and applications, *Electr. Pow. Syst. Res.*, vol. 214, p. 108725, 2023.

[11] J. Kazempour, P. Pinson, and B. F. Hobbs, "A stochastic market design with revenue adequacy and cost recovery by scenario: Benefits and costs," *IEEE Trans. Power Syst.*, vol. 33, no. 4, pp. 3531–3545, 2018.

[12] W. B. Powell and S. Meisel, "Tutorial on stochastic optimization in energy—Part I: Modeling and policies," *IEEE Trans. Power Syst.*, vol. 31, no. 2, pp. 1459–1467, 2015.

[13] N. Aguiar, V. Gupta, and P. P. Khargonekar, "A real options market-based approach to increase penetration of renewables," *IEEE Trans. Smart Grid*, vol. 11, no. 2, pp. 1691–1701, 2020.

[14] R. Chen, A. Botterud, H. Sun, and Y. Wang, "A bilateral reserve market for variable generation: Concept and implementation," in *2016 IEEE Power and Energy Society General Meeting (PESGM)*, 2016, pp. 1–5.

[15] FERC Staff Paper, "Energy and ancillary services market reforms to address changing system needs," Federal Energy Regulatory Commission (FERC), Docket NO. AD21-10- 000, 2021.

[16] H. Shin and R. Baldick, "Mitigating market risk for wind power providers via financial risk exchange," *Energy Econ.*, vol. 71, pp. 344–358, 2018.

[17] C. Johnathon, A. P. Agalgaonkar, C. Planiden, and J. Kennedy, "A proposed hedge-based energy market model to manage renewable intermittency," *Renew. Energy*, vol. 207, pp. 376–384, 2023.

[18] B. Frew, G. Brinkman, *et al.*, "Impact of operating reserve rules on electricity prices with high penetrations of renewable energy," *Energy Policy*, vol. 156, p. 112443, 2021.

[19] J. Cartuyvels and A. Papavasiliou, "Calibration of operating reserve demand curves using a system operation simulator," *IEEE Trans. Power Syst.*, vol. 38, no. 4, pp. 3043–3055, 2023.

[20] K. Bruninx and E. Delarue, "Endogenous probabilistic reserve sizing and allocation in unit commitment models: Cost-effective, reliable, and fast," *IEEE Trans. Power Syst.*, vol. 32, no. 4, pp. 2593–2603, 2017.

[21] W. W. Hogan and S. L. Pope, "PJM Reserve Markets: Operating reserve demand curve enhancements," 2019, Center for Business and Government, JFK School of Government, Harvard University, Cambridge, MA.

[22] J. Mays, "Quasi-stochastic electricity markets," *INFORMS Journal on Optimization*, vol. 3, no. 4, pp. 350–372, 2021.

[23] California Independent System Operator, "Day-ahead Market Enhancements," Draft Final Proposal, Dec 1 2022.

[24] K. Wilker and A. Zhou, "Methodologies to determine IRU eligibility price cap, Day-ahead Market Enhancements," Dec 1 2022, CAISO.

[25] "Comments on draft final proposal day-ahead market enhancements." [Online]. Available: https://stakeholdercenter.caiso.com/Comments/AllComments/9ad17b1a-1975-490c-85f0-c52d3984e28e

[26] California Independent System Operator, "Day-ahead Market Enhancements," Final Proposal, January 11, 2023.

[27] M. G. Pollitt and K. L. Anaya, "Competition in markets for ancillary services? the implications of rising distributed generation," Energy J., vol. 41, 2020, Special issue 1: Competition in the Electricity Sector.

[28] X. Fang, H. Cui, *et al.*, "Characteristics of locational uncertainty marginal price for correlated uncertainties of variable renewable generation and demands," *Appl. Energy*, vol. 282, p. 116064, 2021.

[29] S. Chang, L. Xie, and J. Dumas, "A reserve forecast-based approach to determining credit collateral requirements in electricity markets," in *2015 IEEE Power & Energy Society General Meeting*, 2015, pp. 1–5.

[30] UL, Renewable Energy Forecasting for Energy Integration, [Online]. Available: https://www.ul.com/software/renewable-energy-forecasting-energy-integration.

[31] L. Matsunobu, "Adapting probabilistic forecasts based on conditions," in *2024 ESIG Forecasting and Markets Workshop*, Salt Lake City, UT, USA. 2024. [Online]. Available: https://www.esig.energy/download/session-5b-adapting-probabilistic-forecasts-based-on-conditions-lysha-matsunobu/?wpdmdl=11686&refresh=666c647e264611718379646

[32] H.-P. Chao and R. Wilson, "Priority service: Pricing, investment, and market organization," *Am. Ec. Rev.*, vol. 77, no. 5, pp. 899–916, 1987.

[33] Https://github.com/elinaspyrou/Flexibility_Options_Numerical_Study.

[34] K. Moyer and D. Ramirez, "EDAM Benefits Study Estimating Savings for California and the West Under EDAM Market Scenarios", Prepared for CAISO by Energy Strategies, Nov 2022. [Online]. Available: https://www.caiso.com/Documents/Presentation-CAISO-Extended-Day-Ahead-Market-Benefits-Study.pdf.